\documentclass[11pt,journal,onecolumn]{IEEEtran}

\usepackage{cite}
\usepackage{graphicx}
\usepackage{booktabs}
\usepackage{multirow}
\usepackage{array}
\usepackage{xcolor}
\usepackage{colortbl}
\usepackage{caption}
\usepackage{makecell}
\usepackage{setspace}
\usepackage{textcomp}
\usepackage{parskip}
\usepackage{microtype}
\usepackage{hyperref}

\definecolor{headerblue}{RGB}{46,117,182}
\definecolor{rowgray}{RGB}{238,244,251}
\definecolor{notecolor}{RGB}{90,90,90}

\newcolumntype{L}[1]{>{\raggedright\arraybackslash}p{#1}}
\newcolumntype{C}[1]{>{\centering\arraybackslash}p{#1}}

\title{Cross-Modal Fusion of OCT and OCT angiography enface for Improved Diagnostics of Diabetic Retinopathy}

\author{
Rashadul Hasan Badhon,
Atalie Carina Thompson,
Jennifer I. Lim,
Theodore Leng,
and Minhaj Nur Alam%
}

\begin{document}

\maketitle
\section{Abstract}
Diabetic retinopathy (DR) is a leading cause of vision impairment worldwide, highlighting the need for accurate and accessible screening tools. Optical Coherence Tomography (OCT) provides high-resolution structural information of the retina, whereas OCT angiography (OCTA) offers complementary vascular information that is highly relevant for DR diagnosis. In this study, we propose a cross-modal fusion of OCT B-scans with single-channel enface OCTA using a bidirectional cross-modal attention network for automated DR classification. Two independent datasets, OCT500 and UIC, comprising 730 subjects in total, were utilized to evaluate performance under within-dataset, combined-dataset and cross-dataset generalization settings. A ConvNeXt V2 model trained solely on OCT images served as the unimodal baseline. In addition to ground-truth (GT) OCTA, we explored the use of translated (TR) OCTA generated from OCT scans, eliminating the requirement for dedicated OCTA hardware. Experimental results demonstrate that cross-modal fusion consistently outperforms unimodal OCT classification across all evaluation scenarios. Fusion with GT OCTA improved classification accuracy and discriminative performance, while TR OCTA achieved comparable or superior results in most settings. Furthermore, TR OCTA improved sensitivity and cross-dataset generalization, indicating enhanced robustness to domain shifts. These findings demonstrate that attention-based OCT–OCTA enface fusion provides clinically meaningful improvements for DR detection and suggest that computationally generated OCTA can serve as a practical, low-cost alternative to hardware-acquired OCTA, enabling broader deployment of high-performance retinal screening systems in resource-limited clinical environments.



\section{Introduction}

Optical Coherence Tomography (OCT) is a non-invasive high-resolution imaging modality, using low-coherence interferometry to produce depth-resolved cross-sectional images of the microstructure of biological tissue with micrometer-scale resolution \cite{huang1991optical,fercher2003optical}. Since its initial demonstration, OCT has become popular in medical imaging due to its ability to visualize subsurface morphology in real time without contact or any invasive medical procedure. In the field of ophthalmology, OCT has revolutionized the assessment of retinal architecture and pathologies, enabling precise evaluation of retinal layers, photoreceptor integrity and subtle structural changes associated with major causes of vision loss such as age-related macular degeneration (AMD), diabetic retinopathy (DR) and glaucoma \cite{drexler2008optical,schmidt2018retina}. \\Although traditional OCT captures structural retinal information, OCT Angiography (OCTA) extends OCT to visualize vascular networks non-invasively by detecting motion contrast between sequential B-scans at the same spatial location, thus highlighting blood flow dynamics without the need to introduce contrast agents \cite{wang2017optical}. OCTA provides depth-resolved en face projections of retinal and choroidal microvasculature, offering new quantitative biomarkers such as vessel density, caliber etc which are critical for early detection and longitudinal monitoring of retinal vascular diseases \cite{gao2016quantitative,spaide2018optical}.
The enface representation of OCTA data, projecting volumetric information onto a plane parallel to the retinal surface, is especially useful for visualizing top-down vascular maps and broader spatial patterns hidden within individual B-scans \cite{an2016quantitative}. OCTA enface, hence facilitates visualization of the vascular architecture in a global context, while individual OCT B-scans provide high-resolution structural contrast across depth. \\
Most of the commercially available devices produce OCT slices and OCTA enface. Not to mention that OCTA requires additional expensive hardware which is not always affordable or available at the local clinics. Hence, for retinal pathology detection, OCT has been widely utilized so far, despite the clear merits of OCTA. Given the additional vascular information present in OCTA, combining it with the already available OCT, disease prognosis and progression monitoring would be faster and more accurate, leading to better medical services. Thus cross-modal fusion comes to the scenario.\\
Recently, cross-modal retrieval has gained popularity due to the rapid increase in multimodal data thanks to the fast technological development in medical science. The idea of high level semantic homogeneity and low level experssive heterogeneity brought forth the necessity of combining multiple forms of data into one for better representation. Unimodal data handles information from one channel only, limiting any utilization from other modality data, thereby degrading potential performance improvement. Cross-modal image fusion has thus emerged as a pivotal paradigm in clinical diagnosis and research, as single-modality images often provide only fragmented representations of complex disease processes \cite{li2024review}. This provides an opportunity to integrate complementary information from different imaging modalities to provide a more comprehensive understanding of the underlying pathology, leading to improved diagnostic accuracy and treatment planning \cite{azam2022review,haribabu2023recent}. Recent studies have demonstrated that deep learning-based multimodal fusion techniques have become powerful tools for medical image detection and prognosis \cite{li2024review}. \\In general, fusion approaches can be broadly categorized into three main schemes: early fusion (inputl level), intermediate fusion (feature-level) and late fusion (output level)\cite{li2024review,pei2023review}.
Early fusion combines raw data from different modalities at the input level, allowing the network to learn joint representations from the outset. However, this approach may struggle to identify meaningful correlations between modalities when input characteristics differ substantially in terms of intensity, spatial patterns or noise profiles \cite{ryu2023optimizing}. Late fusion, on the other hand, combines predictions from modality-specific networks at the decision stage, offering flexibility but potentially sacrificing opportunities for deep feature interaction \cite{heisler2020ensemble}. Intermediate fusion, however, merges features at various depths within the network architecture and has demonstrated superior performance by enabling the model to learn hierarchical cross-modal correlations at different levels of abstraction \cite{he2021multimodal,ryu2023optimizing}.\\
Recent advances have introduced attention mechanisms and Transformer architectures to address the limitations of traditional convolutional neural network (CNN) based fusion methods. CNNs excel at extracting local features through convolutional operations but have limited receptive fields that hinder their ability to capture long-range dependencies and global contextual information—essential for accurately integrating diverse modalities \cite{sun2024cmaf,li2024crossfuse}. Vision transformers (ViTs) leverage multi-head self-attention mechanisms to model global context and long-distance relationships, proving highly effective in medical image fusion tasks \cite{qu2022transmef,tang2022matr}. Hybrid architectures combining CNNs and transformers have emerged as particularly promising, exploiting CNNs' local feature extraction capabilities alongside transformers' global modeling strengths \cite{wang2024mactfusion,yang2024ecfusion}. Cross-attention mechanisms that are specifically designed for multimodal fusion have been proposed in recent studies to enhance complementary information extraction between modalities while suppressing redundant features \cite{li2024crossfuse}.\\
In the context of neuroimaging, multimodal fusion of structural MRI and functional MRI using cross-attention mechanisms has significantly improved diagnostic accuracy for brain disorders \cite{liu2024multivit}. Similarly, in oncology, multimodal deep learning fusion combining radiological images with histopathological data has enhanced tumor classification and grading performance \cite{huang2024comprehensive}. These successes underscore the potential of sophisticated fusion strategies to leverage complementary relationships among clinical modalities for improved disease characterization and prognostic modeling \cite{huang2024comprehensive}.
In ophthalmology, the integration of structural OCT and vascular OCTA information represents a natural and clinically meaningful fusion paradigm. Structural OCT provides detailed morphological information about retinal layers and tissue architecture, while OCTA reveals functional blood flow patterns and microvascular perfusion. Several studies have demonstrated that combining these complementary modalities improves classification performance for retinal diseases. For instance, multimodal training processes incorporating both enface OCT and OCTA have been successfully employed for automated artery-vein classification, where enface OCT provides vessel intensity profiles similar to near-infrared fundus images and OCTA contributes blood flow strength and vessel geometry features \cite{alam2020avnet,abtahi2022mfavnet}. \\
Some recently published works have systematically evaluated different fusion strategies for OCTA layer combinations in retinal pathology classification. Intermediate fusion architectures, which merge features from multiple OCTA layers (superficial capillary plexus, deep capillary plexus, and choriocapillaris) at intermediate network depths, have achieved superior performance compared to early or late fusion approaches \cite{ryu2023optimizing}. The intermediate fusion strategy enables the model to learn and exploit complementary information from different vascular layers at appropriate levels of feature abstraction, addressing the limitation that early fusion may not allow CNNs to discriminate layer-specific pathological patterns effectively \cite{ryu2023optimizing}. Hierarchical fusion approaches combining structural OCT and OCTA with additional imaging modalities have also demonstrated advantages over simpler fusion schemes in DR detection \cite{liu2023hybrid}. A recent study \cite{Hongyi Pan 2025} showed that OCT coupled with multiple layer-specific projections feeding into a simple CNN can result in a better performance compared to early or late fusion. \\
However, challenges remain in effectively fusing the OCT and OCTA modalities that are available from commercial devices, since in a clinical settings the OCTA enface is often the only data available in addition to the usual OCT slices. Data heterogeneity resulting from modality-specific noise, resolution variations, and inconsistent annotations can impede performance and generalizability  hence we need more than a simple CNN to process and extract the multimodal information. The requirement for paired multimodal data during both training and inference stages limits the broader adoption of fusion methods, as large-scale paired datasets are difficult to acquire \cite{liu2024multieye}. Moreover, maintaining significant cross-modal correlations during fusion while ensuring interpretability is crucial for clinical implementation hence adoption remains an open challenge \cite{huang2024comprehensive}.
In parallel with advances in hardware and fusion methodologies, deep learning and computational methods have been increasingly applied to OCT and OCTA imaging to enable automated classification, segmentation, and synthetic image generation \cite{ran2019development,li2019deep}. Deep learning models have demonstrated the feasibility of synthesizing OCTA-like vascular maps from structural OCT scans by learning reflectance–flow correlations \cite{Srinivasan2021}. Building upon these foundations, our work investigates cross-modal fusion of structural OCT and enface OCTA for improved DR classification. Our novelty is several folds: (1) single channel enface OCTA is used as the 2nd modality to mimic a real life clinical settings of scarce data availability, (2) a proper Cross modal attention framework adapted for feature extraction and mixing, (3) We used 2 different datasets in different training arrangements to validate the performance against a based model classifier and (4) Use of translated (TR) OCTA enface replacing the ground truth (GT) OCTA enface generated from the OCT slices. This is significant from the perspective of affordable healthcare where instead of additional hardware and expenditure, we can utilize the already available dataset and get better performance. We have generated the TR OCTA enface after translating OCTA slices from the corresponding OCT slices using a conditional diffusion model, trained on OCT500 only \cite{diffusion_mine}.\\
Across our different training configurations, we found that the fusion of OCT and OCTA enface significantly improves performance over an unimodal model in terms of accuracy, precision, recall, F1, AUC and specificity. We also included confidence interval to validate the fusion process and generated results. On top of that we introduced TR OCTA as an alternative to acquire the 2nd modality data and the performance improved in most cases compared to the GT OCTA data.\\
The primary contributions of this work are fourfold. First, we demonstrated that cross-modal attention-based fusion of OCT  with a single-channel OCTA enface consistently and substantially outperforms a strong unimodal ConvNeXt V2 baseline for DR classification across multiple within-dataset and cross-dataset scenarios. Second, we utilized TR OCTA enface, translated entirely from OCT slices via a conditional diffusion model, removing the dependency on physical OCTA hardware and this TR OCTA fusion variant matches or exceeds GT OCTA fusion performance in the majority of experimental configurations. They also achieved higher sensitivity in cross-dataset generalization, making high-quality multimodal classification accessible with only a standard OCT scanner. Third, we provide a rigorous multi-scenario evaluation framework: spanning single or combined dataset training and bidirectional cross-dataset generalization across two independent cohorts (OCT500 and UIC) with distinct acquisition protocols and patient demographics. This was accompanied by 95\% bootstrap confidence intervals, giving a statistically grounded assessment of both performance and uncertainty for this study. Fourth, by showing that a computationally synthesized second modality can substitute for expensive OCTA hardware without sacrificing diagnostic performance, this work establishes a clinically meaningful pipeline for affordable, high-sensitivity DR screening in resource-limited settings where OCTA devices are unavailable, offering a practical bridge between well-resourced specialist clinics and underserved primary care environments.\\

\section{Materials and Methods}
\subsection{Dataset}
\begin{figure}[htbp]
    \centering
    \includegraphics[width=0.5\linewidth]{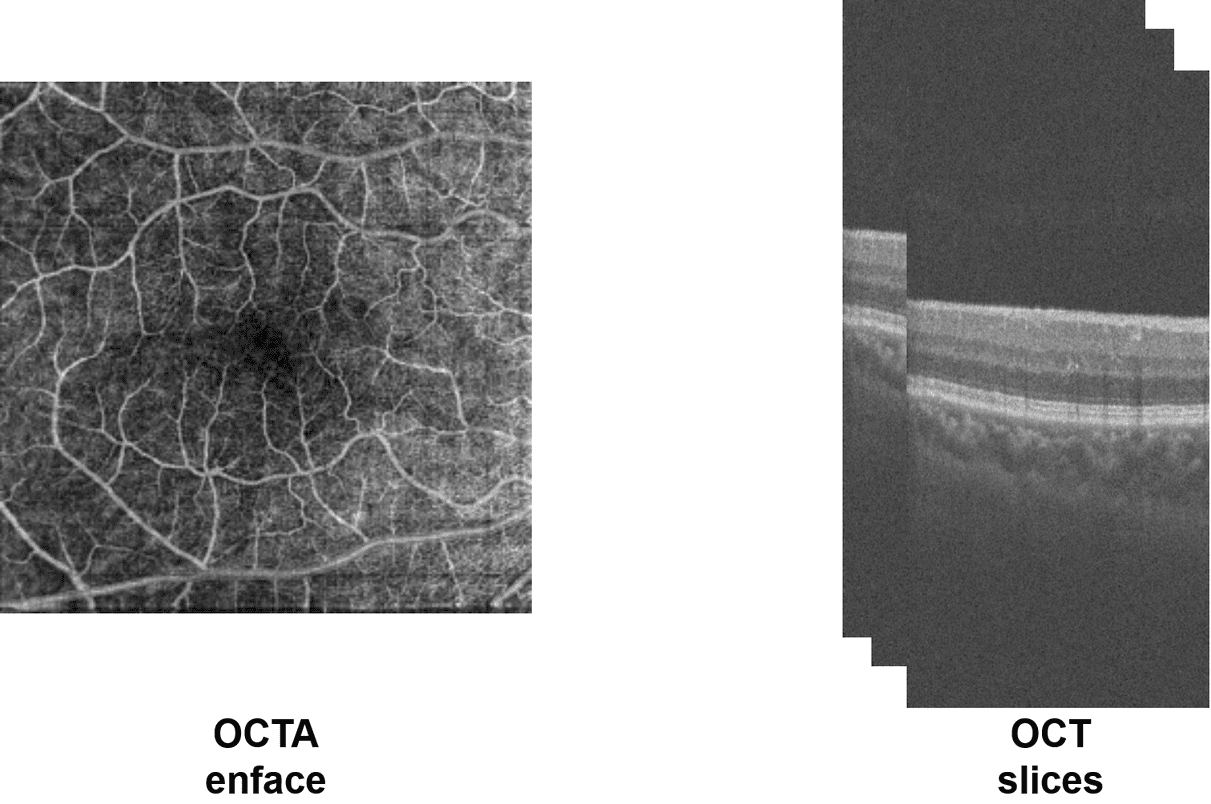}
    \caption{OCT and OCTA enface sample data}
    \label{fig:dataset}
\end{figure}

In this study, two independent retinal imaging datasets were used: OCT500 \cite{OCT500dataset} and UIC.
OCT500 is a publicly available dataset comprising OCT B-scan volumes and corresponding
enface OCTA images acquired for 3mm and 6mm field of views (FoV). Out of 500 patients only 251 normal and
64 DR patients were used for this study since UIC has only normal and DR labels.
UIC is another dataset collected at the University of Illinois Chicago, containing paired OCT
and OCTA acquisitions from 64 normal and 351 DR patients. Together, the two datasets
provide complementary sources of multi-modal retinal data that differ in acquisition
device, imaging protocol and patient demographics, yielding a combined pool of 315
normal and 415 DR patients (730 total). For each patient, unimodal input consists of only OCT slices. On the other hand, the multi-modal input consists of one enface OCTA image and approximately 80 OCT B-scan slices (Fig.~\ref{fig:dataset}) centred around the foveal region, giving a rich
cross-sectional view of the macular tissue alongside the corresponding vascular
projection. We also compare performance for TR OCTA enface replacing GT OCTA enface as an alternative, where we can utilize OCTA translation from OCT to enhance multimodal fusion scope. Stratified splits were applied at the Patient-level to prevent data leakage
across partitions.\\
\\
To systematically evaluate both the performance and the generalisation capability of the proposed models, five training and evaluation scenarios were planned. In the first two scenarios, each dataset was used independently: models were trained and tested exclusively on OCT500 or UIC data using an 80-10-10\%\ train-validation-test split to establish dataset-specific performance benchmarks. In the third scenario, the two datasets were pooled into a single combined dataset, and the model was trained and tested on this merged collection, assessing whether a larger and more heterogeneous training set improves overall performance. The final two scenarios were designed to evaluate cross-dataset generalisation: in one, the model was trained on OCT500 (90-10\% split) and tested on UIC; in the other, the training and test sets were reversed, with UIC used for training (90-10\% split) and OCT500 for testing. These cross-dataset experiments probe how well models trained on one imaging environment transfer to an unseen acquisition setting, which is a critical consideration for clinical deployment.

\subsection{Model framework}
\begin{figure}[htbp]
    \centering
    \includegraphics[width=0.8\linewidth]{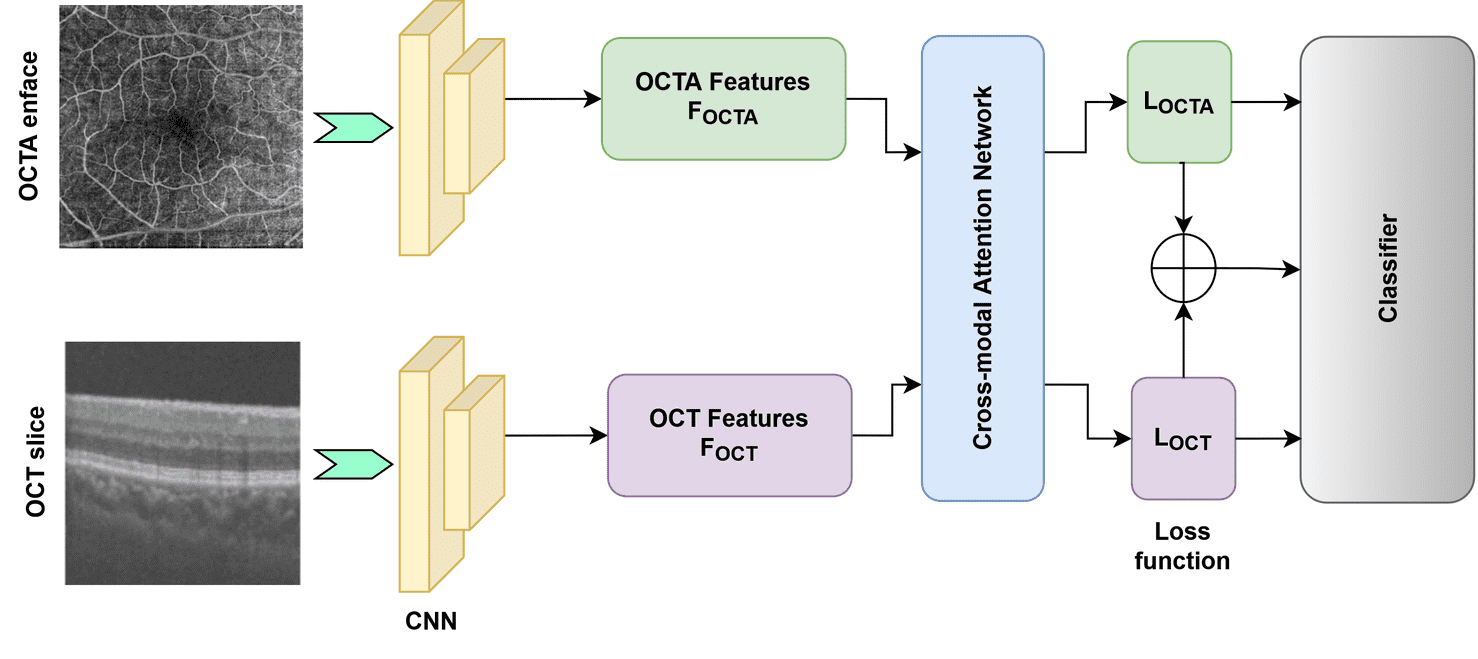}
    \caption{Cross-modal Fusion framework}
    \label{fig:crdnet}
\end{figure}

To establish a single-modality baseline for comparison against the multi-modal fusion approach, we trained a ConvNeXt Large network \cite{convnextv2} independently on OCT B-scan images. The model was initialised with ImageNet-1k pretrained weights and its final classification head was replaced with a linear layer sized to the number of target classes. The network was trained end-to-end with a learning rate of $1\times10^{-4}$. Input images were resized to 224×224 pixels and normalised, with no modality fusion or cross-modal attention applied. Performance was evaluated across the five scenarios mentioned above using accuracy, precision, recall, F1 score, AUC and specificity, while using the same data splits and evaluation protocol as CRD-Net, enabling direct comparison between single-modal and multi-modal strategies.\\
For the multimodal fusion model architecture, we adopted the Cross-modal Retinal Disease Diagnosis Network (CRD-Net) proposed by Liu et al. \cite{mainreference} as the fusion backbone to combine OCT B-scan and enface OCTA modalities. CRD-Net (Fig.~\ref{fig:crdnet}) is designed to explicitly model the spatial correspondence between paired multi-modal retinal images, rather than treating each modality independently.\\
\\
The architecture consists of three principal components: (1) feature extraction via independent convolutional neural networks (CNNs), (2) a cross-modal Attention module that enables bidirectional feature exchange between modalities and (3) a multi-branch classifier trained with a composite loss function.\\
Each input modality: the OCT B-scan and the en face OCTA image, is processed by a separate CNN backbone to extract modality-specific instance-level feature representations. From the two modalities, two feature maps are extracted. The two backbones share the same architecture but maintain independent weights, allowing each stream to learn representations tailored to its modality's characteristics. For all the experiments, we used ImageNet-pretrained weights as initialization. The final feature representations are taken from the last layer of each backbone, providing globally pooled, semantically rich feature maps prior to the fusion.\\
The attention module is motivated by the clinical workflow in which ophthalmologists identify a lesion in one imaging modality and then query the corresponding anatomical location in the other modality to corroborate their findings. This module consists of two sequential sub-components: a specific modal attention module and a cross-modal fusion module.\\
Within each stream, a multi-head self-attention mechanism is first applied to the feature maps to direct the network’s attention toward disease-relevant spatial regions within that modality. Specifically, each feature map is first projected via a 1×1 convolution, passed through the attention network to produce an intermediate feature and then further refined by a feed forward network with a residual connection back to the original feature map. This yields modality-specific attended outputs.\\
After per-modality attention, the fusion module performs bidirectional cross-modal querying. In the OCT branch, the attended OCT features are projected to form the key and value matrices, while the attended OCTA features serve as the query. A scaled dot-product attention is computed between these projections, and the result is added back via a residual connection to produce the OCT branch output. This allows the OCT branch to identify which of its own regions are most relevant to the lesion patterns observed in the enface OCTA. The symmetric operation is applied in the OCTA branch in a similar way to yield the OCTA branch output. Residual connections are used throughout to support stable gradient flow. The attention module is inserted at the output of the final layer as it ensures that the input features are sufficiently abstract and semantically rich before cross-modal interaction is performed. Following the attention module, the cross-attended feature maps are concatenated to form a unified multi-modal representation, which is passed to a fully connected classification head. Two additional single-modality classifiers are simultaneously trained on each stream’s attended features independently. The total training objective is the sum of three cross-entropy losses: one from the fused multi-modal branch, one from the OCT-only branch, and one from the OCTA-only branch, all jointly optimised via Stochastic Gradient Descent (SGD). This composite loss ensures balanced supervision across both modalities, discouraging over-reliance on either single stream and reducing modality-specific bias during training.\\
All models were implemented in PyTorch. Following the original CRD-Net configuration, the optimiser was SGD with momentum 0.9 and weight decay $1\times10^{-4}$. The initial learning rate was set to $1\times10^{-4}$ and decayed by a factor of 0.1 at epochs 50 and 100, training for a total of 100 epochs with a batch size of 32. Input images from both modalities were augmented identically using random horizontal flipping (probability 0.5) and random rotation within ±30°. All backbone weights were initialised from ImageNet pre-trained checkpoints.

\section{Results}

We compared the classification performance of both the unimodal baseline and the proposed
cross-modal fusion model across three within-dataset training configurations: OCT500, UIC and
Combined and two cross-dataset generalization scenarios: OCT500-trained model tested on UIC, and
UIC-trained model tested on OCT500. We reported performance using accuracy, precision, recall,
F1-score, AUC and specificity. For the cross-modal fusion model, all metrics are accompanied by
95\% bootstrap confidence intervals (CIs) computed over 1{,}000 resampling iterations to provide
reliable estimates of statistical uncertainty.

\subsection*{3.1~Unimodal Baseline Performance}

Table~\ref{tab:unimodal} represents the classification performance of the unimodal ConvNeXt V2
baseline model trained solely on OCT images. When trained and tested within the same dataset, the
model achieves reasonable performance: an accuracy of 0.8634 on OCT500 and 0.9621 on UIC.
The combined dataset yields an accuracy of 0.9221 with balanced recall (0.9278) and F1-score
(0.9215).

However, cross-dataset generalization reveals a significant performance drop. When trained on
OCT500 and tested on UIC, accuracy falls to 0.5735 with an F1-score of only 0.5396. UIC trained model when tested on OCT500 improves slightly to an accuracy of 0.7048, but the F1-score remains limited at
0.6456.

\begin{table}[ht]
\centering
\caption{Unimodal baseline (ConvNeXt V2) classification performance}
\label{tab:unimodal}
\renewcommand{\arraystretch}{1.3}
\begin{tabular}{|L{2.5 cm}|C{1.75cm}|C{1.75cm}|C{1.75cm}|C{1.75cm}|C{1.75cm}|C{1.75 cm}|}
\hline
\textbf{Training Arrangement} &
\textbf{Accuracy} &
\textbf{Precision} &
\textbf{Recall} &
\textbf{F1-Score} &
\textbf{AUC} &
\textbf{Specificity} \\
\hline
OCT500                            & 0.8634 & 0.8977 & 0.6872 & 0.7309 & 0.7994 & 0.6872 \\
\hline
UIC                               & 0.9621 & 0.9376 & 0.9212 & 0.9292 & 0.9882 & 0.9212 \\
\hline
Combined                          & 0.9221 & 0.9204 & 0.9278 & 0.9215 & 0.9764 & 0.9278 \\
\hline
OCT500 train UIC test     & 0.5735 & 0.6278 & 0.7396 & 0.5396 & 0.9003 & 0.7396 \\
\hline
UIC train OCT500 test     & 0.7048 & 0.6456 & 0.7136 & 0.6456 & 0.7939 & 0.7136 \\
\hline
\end{tabular}
\end{table}

\subsection*{3.2~Cross-Modal Fusion Performance with GT OCTA }

Table~\ref{tab:gt_octa} presents the performance of the cross-modal fusion model using
GT OCTA enface images. Within-dataset performance is substantially improved over
the unimodal baseline. On OCT500, the fusion model achieves an accuracy of 0.8788
(95\%~CI:~0.8646--0.8896) and an AUC of 0.9719 (95\%~CI:~0.9656--0.9774). On UIC, performance is
near-ceiling with an accuracy of 0.9971 (95\%~CI:~0.9954--0.9989) and an AUC of 1.0000.
The combined dataset yields an accuracy of 0.9730 (95\%~CI:~0.9688--0.9768) with well-balanced
precision and recall.

In the cross-dataset generalization, the OCT500-trained model tested on UIC, achieves an
accuracy of 0.8746 (95\%~CI:~0.8713--0.8783) and an AUC of 0.9651 (95\%~CI:~0.9631--0.9671),
a notable improvement over the unimodal baseline. However, recall drops to
0.5935 (95\%~CI:~0.5881--0.5988). The UIC-trained, OCT500-tested case
yields an accuracy of 0.8345 (95\%~CI:~0.8297--0.8386) and an AUC of 0.7216
(95\%~CI:~0.7137--0.7299), suggesting a stronger domain gap when generalizing from the richer
UIC acquisition protocol to OCT500.

\begin{table}[ht]
\centering
\caption{Cross-modal fusion performance (GT OCTA) with 95\% bootstrap confidence
intervals}
\label{tab:gt_octa}
\renewcommand{\arraystretch}{1.5}
\begin{tabular}{|L{2.5 cm}|C{1.75cm}|C{1.75cm}|C{1.75cm}|C{1.75cm}|C{1.75cm}|C{1.75 cm}|}
\hline
\textbf{Training Arrangement} &
\textbf{Accuracy} &
\textbf{Precision} &
\textbf{Recall} &
\textbf{F1-Score} &
\textbf{AUC} &
\textbf{Specificity} \\
\hline
OCT500 &
  \makecell{0.8788 \\ {\scriptsize(0.8646--0.8896)}} &
  \makecell{0.9333 \\ {\scriptsize(0.9257--0.9395)}} &
  \makecell{0.7143 \\ {\scriptsize(0.6935--0.7355)}} &
  \makecell{0.7643 \\ {\scriptsize(0.7406--0.7871)}} &
  \makecell{0.9719 \\ {\scriptsize(0.9656--0.9774)}} &
  \makecell{0.7143 \\ {\scriptsize(0.6935--0.7355)}} \\
\hline
UIC &
  \makecell{0.9971 \\ {\scriptsize(0.9954--0.9989)}} &
  \makecell{0.9983 \\ {\scriptsize(0.9973--0.9993)}} &
  \makecell{0.9912 \\ {\scriptsize(0.9856--0.9964)}} &
  \makecell{0.9947 \\ {\scriptsize(0.9914--0.9979)}} &
  \makecell{1.0000 \\ {\scriptsize(1.0000--1.0000)}} &
  \makecell{0.9912 \\ {\scriptsize(0.9856--0.9964)}} \\
\hline
Combined &
  \makecell{0.9730 \\ {\scriptsize(0.9688--0.9768)}} &
  \makecell{0.9706 \\ {\scriptsize(0.9662--0.9748)}} &
  \makecell{0.9762 \\ {\scriptsize(0.9727--0.9795)}} &
  \makecell{0.9727 \\ {\scriptsize(0.9685--0.9766)}} &
  \makecell{0.9944 \\ {\scriptsize(0.9932--0.9953)}} &
  \makecell{0.9762 \\ {\scriptsize(0.9727--0.9795)}} \\
\hline
OCT500 train UIC test &
  \makecell{0.8746 \\ {\scriptsize(0.8713--0.8783)}} &
  \makecell{0.9354 \\ {\scriptsize(0.9337--0.9373)}} &
  \makecell{0.5935 \\ {\scriptsize(0.5881--0.5988)}} &
  \makecell{0.6230 \\ {\scriptsize(0.6149--0.6310)}} &
  \makecell{0.9651 \\ {\scriptsize(0.9631--0.9671)}} &
  \makecell{0.5935 \\ {\scriptsize(0.5881--0.5988)}} \\
\hline
UIC train OCT500 test &
  \makecell{0.8345 \\ {\scriptsize(0.8297--0.8386)}} &
  \makecell{0.8301 \\ {\scriptsize(0.8200--0.8401)}} &
  \makecell{0.6111 \\ {\scriptsize(0.6056--0.6171)}} &
  \makecell{0.6353 \\ {\scriptsize(0.6275--0.6436)}} &
  \makecell{0.7216 \\ {\scriptsize(0.7137--0.7299)}} &
  \makecell{0.6111 \\ {\scriptsize(0.6056--0.6171)}} \\
\hline
\multicolumn{7}{l}{\scriptsize Values in parentheses represent 95\% bootstrap confidence intervals (1{,}000 iterations).}
\end{tabular}
\end{table}

\subsection*{3.3~Cross-Modal Fusion Performance with TR OCTA }

Table~\ref{tab:tr_octa} presents the cross modal fusion performance when TR OCTA enface is utilized in stead of GT OCTA. Within-dataset performance is markedly higher than both the unimodal baseline
and the GT OCTA fusion variant. On OCT500, the model achieves an accuracy of 0.9697
(95\%~CI:~0.9630--0.9761) and an AUC of 0.9897 (95\%~CI:~0.9862--0.9931), with recall of
0.9286 (95\%~CI:~0.9140--0.9432). On UIC, performance is similarly higher than the unimodal base model.
The combined dataset achieves the strongest overall metrics: accuracy 0.9820
(95\%~CI:~0.9786--0.9853), recall 0.9792 (95\%~CI:~0.9754--0.9829), and AUC 0.9991
(95\%~CI:~0.9988--0.9994), demonstrating that TR OCTA provides more consistent and informative cross-modal features for classification.

Cross-dataset generalization is also improved under TR OCTA compared to the base model. The OCT500-trained model tested on
UIC achieves an accuracy of 0.8829 (95\%~CI:~0.8795--0.8862) and a recall of 0.9013
(95\%~CI:~0.8976--0.9049), indicating substantially better sensitivity compared to the GT OCTA
counterpart. The UIC-trained model tested on OCT500 yields an accuracy of 0.7005
(95\%~CI:~0.6950--0.7057) and an AUC of 0.8047 (95\%~CI:~0.7989--0.8110). While this remains
the most challenging scenario, the TR OCTA fusion model still outperforms the unimodal baseline.

\begin{table}[ht]
\centering
\caption{Cross-modal fusion performance (TR OCTA) with 95\% bootstrap confidence
intervals}
\label{tab:tr_octa}
\renewcommand{\arraystretch}{1.5}
\begin{tabular}{|L{2.5 cm}|C{1.75cm}|C{1.75cm}|C{1.75cm}|C{1.75cm}|C{1.75cm}|C{1.75 cm}|}
\hline
\textbf{Training Arrangement} &
\textbf{Accuracy} &
\textbf{Precision} &
\textbf{Recall} &
\textbf{F1-Score} &
\textbf{AUC} &
\textbf{Specificity} \\
\hline
OCT500 &
  \makecell{0.9697 \\ {\scriptsize(0.9630--0.9761)}} &
  \makecell{0.9815 \\ {\scriptsize(0.9774--0.9853)}} &
  \makecell{0.9286 \\ {\scriptsize(0.9140--0.9432)}} &
  \makecell{0.9521 \\ {\scriptsize(0.9415--0.9623)}} &
  \makecell{0.9897 \\ {\scriptsize(0.9862--0.9931)}} &
  \makecell{0.9286 \\ {\scriptsize(0.9140--0.9432)}} \\
\hline
UIC &
  \makecell{0.9699 \\ {\scriptsize(0.9638--0.9753)}} &
  \makecell{0.9826 \\ {\scriptsize(0.9793--0.9857)}} &
  \makecell{0.9074 \\ {\scriptsize(0.8922--0.9223)}} &
  \makecell{0.9401 \\ {\scriptsize(0.9290--0.9505)}} &
  \makecell{1.0000 \\ {\scriptsize(1.0000--1.0000)}} &
  \makecell{0.9074 \\ {\scriptsize(0.8922--0.9223)}} \\
\hline
Combined &
  \makecell{0.9820 \\ {\scriptsize(0.9786--0.9853)}} &
  \makecell{0.9845 \\ {\scriptsize(0.9817--0.9874)}} &
  \makecell{0.9792 \\ {\scriptsize(0.9754--0.9829)}} &
  \makecell{0.9816 \\ {\scriptsize(0.9781--0.9849)}} &
  \makecell{0.9991 \\ {\scriptsize(0.9988--0.9994)}} &
  \makecell{0.9792 \\ {\scriptsize(0.9754--0.9829)}} \\
\hline
OCT500 train UIC test &
  \makecell{0.8829 \\ {\scriptsize(0.8795--0.8862)}} &
  \makecell{0.7798 \\ {\scriptsize(0.7746--0.7852)}} &
  \makecell{0.9013 \\ {\scriptsize(0.8976--0.9049)}} &
  \makecell{0.8181 \\ {\scriptsize(0.8131--0.8235)}} &
  \makecell{0.9595 \\ {\scriptsize(0.9573--0.9615)}} &
  \makecell{0.9013 \\ {\scriptsize(0.8976--0.9049)}} \\
\hline
UIC train OCT500 test &
  \makecell{0.7005 \\ {\scriptsize(0.6950--0.7057)}} &
  \makecell{0.6551 \\ {\scriptsize(0.6502--0.6600)}} &
  \makecell{0.7328 \\ {\scriptsize(0.7261--0.7390)}} &
  \makecell{0.6498 \\ {\scriptsize(0.6440--0.6557)}} &
  \makecell{0.8047 \\ {\scriptsize(0.7989--0.8110)}} &
  \makecell{0.7328 \\ {\scriptsize(0.7261--0.7390)}} \\
\hline
\multicolumn{7}{l}{\scriptsize Values in parentheses represent 95\% bootstrap confidence intervals (1{,}000 iterations).}
\end{tabular}
\end{table}

Supplemental figure 1-3 depict the confusion matrices and supplemental figure 4-6 represent ROC curves for all the training scenarios where Normal is set as label '0' and DR is set as label '1'. Moreover, supplemental figure 1,4 show base model performance; 2,5 show cross-modal performance fusion pertaining to GT OCTA and 3,6  show cross-modal fusion performance for TR OCTA.\\
 ROC curves of the unimodal baseline model for within-dataset evaluations yield moderate to strong discriminative performance. The UIC and Combined curves rise sharply toward the upper-left corner at low false positive rates, indicating reliable separation of DR from NORMAL cases when training and test distributions are matched. In contrast, the OCT500 curve exhibits a more gradual ascent with visible irregularities. Under cross-dataset conditions, the OCT500-trained model tested on UIC achieves an AUC of 0.9003 despite a notable drop in recall. The UIC-trained model tested on OCT500 yields the lowest AUC of 0.7939, with a curve that departs more slowly from the diagonal, consistent with the greater domain gap when transferring from UIC to OCT500. On the other hand, ROC curves of the cross-modal fusion model with GT OCTA within-dataset performance is substantially stronger than the unimodal baseline. The UIC curve reaches the upper-left corner almost immediately, indicating near-perfect class separation, while the OCT500 and Combined curves both exhibit steep initial ascents before plateauing near maximum sensitivity. In the cross-dataset setting, the OCT500-trained model tested on UIC has the curve climbing sharply at low false positive rates before gradually leveling off. The UIC-trained model tested on OCT500 yields the lowest AUC of 0.7216, with a noticeably shallower curve that tracks closer to the diagonal. Lastly, for TR OCTA enface, across all within-dataset configurations, the curves rise steeply and hug the upper-left boundary, a consistent improvement over the GT OCTA variant. The Combined curve is nearly indistinguishable from a perfect classifier across the full range of thresholds. In the cross-dataset scenarios, the OCT500-trained model tested on UIC achieves an AUC of 0.9595, with a curve shape closely resembling its GT OCTA counterpart but with a marginally higher initial rise, consistent with the improved recall observed in Table III. The UIC-trained model tested on OCT500 achieves an AUC of 0.8047, representing a meaningful improvement over the GT OCTA cross-dataset result of 0.7216; while the curve still departs more gradually from the diagonal than in the within-dataset cases, the overall shape indicates that TR OCTA features generalize more effectively across acquisition domains than GT OCTA features.

\section{Discussion}
In this study, we combined OCT B-scans and single-channel OCTA enface images through a bidirectional cross-modal attention fusion network to classify DR against normal cases. We evaluated performance across five experimental configurations spanning within-dataset, combined dataset and cross-dataset scenarios. Furthermore, we introduced a TR OCTA variant configuration that removes the cost, time consumption and hardware dependency for OCTA data collection. Taken together, the results confirm that cross-modal fusion consistently and substantially outperforms unimodal OCT classification and that the TR OCTA is another potential alternative for improved performance.\\
From Table I,II \&\ III we can observe a consistent performance gap between unimodal and cross modal fusion for all cases. This reflects the fundamental asymmetry in the information content of structural and vascular retinal images. OCT B-scans offer high-resolution depth-resolved views of retinal layer architecture, capturing structural hallmarks of DR such as intraretinal fluid, disruption of the inner segment/outer segment junction, and retinal layer thinning. OCTA enface images, by contrast, reveal the microvascular topology of the inner retina and choriocapillaris, enabling visualization of vascular features: vessel density changes, enlargement of the foveal avascular zone and capillary dropout that emerge early in the DR disease course and are often not directly visible in OCT. We have demonstrated that by combining OCT slices and single-channel enface OCTA, it is possible to learn a richer representation of the disease state that single-modal stream cannot independently achieve, which is evident from the F1-score improvement from the unimodal baseline to the TR OCTA fusion with particularly pronounced gains in recall. Recall is most critical for clinical screening, where false negatives carry the greatest cost to patients.\\
The unimodal base model (Table I) which is trained solely on OCT B-scans, establishes the standard of what structural imaging alone can achieve under our experimental conditions. On the OCT500 dataset, it reaches an accuracy of 0.8634 and an F1-score of 0.7309, while on the UIC dataset it attains higher accuracy and F1 score. For the combined dataset, our base model yields an intermediate accuracy of 0.9221.  We used ConvNeXt V2 for this which serves as a powerful backbone and the results reflect genuine discriminative signal in the OCT B-scan alone. However, the recall on OCT500 sits at only 0.6872, meaning roughly three in ten true DR cases are missed at the fixed operating threshold. This is a direct consequence of the structural ambiguity of early and moderate DR: intraretinal fluid, subtle inner nuclear layer thinning and early photoreceptor disruption are present in OCT, but so are overlapping features from other macular conditions. Without vascular information, the model cannot distinguish DR from Normal patients efficiently.
For comparison, Pan et al. \cite{Hongyi Pan 2025} reported average accuracy of 0.9667, precision of 0.9418, recall of 0.8569, specificity of 0.9422, and F1-score of 0.8921 for their best intermediate fusion model on the OCT500 dataset that include two more retinal conditions in addition to DR and Normal. Our unimodal baseline on the binary DR vs Normal task on OCT500 yields a lower F1-score (0.7309) than Pan et al.’s multi-class model and this discrepancy reflects the different class compositions: their OCT500 test set is drawn from the OCTA-3mm acquisition cohort, whereas our split uses the full OCT500 dataset with a severe 251:64 class imbalance. In a heavily imbalanced binary setting, the model is rewarded for predicting the majority Normal patients, suppressing recall on the minority DR class and lower F1-score accordingly.\\
When trained and tested on separate datasets, our base model performs poorly, which is expected since OCT500 contains mostly Normal patients and UIC is DR dominant dataset. This is most likely because DR produces visually dramatic and consistent pathological characteristics: retinal thickening, fluid accumulation, vascular anomalies, hard exudates. Normal retinas, by contrast, are defined by the absence of pathology, which is a much harder thing to learn. A model trained on UIC sees abundant, feature-rich DR examples and learns a strong positive signal. A model trained on OCT500 has to learn what "normal" looks like from a large pool, while trying to detect subtle early DR from only 64 patients. Lastly, in a combined dataset configuration, the base model works relatively well overall in all performance metrics since the dataset is now close to a balanced state.\\
For cross modal fusion we choose GT OCTA as the second modality and the classification picture changes substantially (Table II). On OCT500, accuracy improves from 0.8634 to 0.8788, but more importantly, AUC jumps from 0.7994 to 0.9719, nearly 18\%\ in discriminative power. On UIC, the model performs extremely well similar to the unimodal case, with accuracy of 0.9971 and AUC of 1.0000, indicating near-perfect separation of DR from Normal patients when the model is trained and tested on the same acquisition protocol. The combined dataset yields high accuracy and AUC, with tightly matched precision and recall, confirming that the fusion benefit scales with dataset size and is not affected by any single cohort’s characteristics. The reason the AUC improves so dramatically even when accuracy improvement is modest on OCT500 is instructive. AUC reflects the model’s ability to rank positive cases above negative ones across all thresholds, while accuracy is measured at a fixed threshold of 0.5. The large AUC gain means that the fusion model has correctly learned that DR patients sit in a very different part of the feature space once vascular information is included, which might limit accuracy for the OCT500 set. The OCTA enface provides the missing pieces needed to complete this picture; the most direct quantitative biomarkers of DR severity and when the cross-modal attention module can query the OCTA branch for these features while processing the OCT stream, it grounds the classification in pathophysiology rather than incidental textural differences. This is consistent with previous studies that attention-mediated fusion of structural and vascular retinal images outperformed concatenation-only approaches, precisely because the attention allows modality-specific pathological signals to be selectively amplified rather than averaged away. If we compare directly with the lates study, our GT OCTA fusion model achieves a relatively lower accuracy of 0.8788 but AUC of 0.9719 compares favorably with their reported AUC for the intermediate fusion model on the same dataset, suggesting that the discriminative power of the two approaches is broadly comparable. To further validate our approach, we used different training configurations and found that multimodal fusion indeed improves performance for all cases. Even cross dataset experiments showed better performance compared to the base model greatly signifying the importance of multimodal information integration for disease detection.\\
Lastly, we replace GT OCTA with TR OCTA, translated entirely from the OCT B-scans via a translation model and found that it not only preserves but consistently keep up with the GT OCTA fusion performance across nearly every metric for almost all experimental configuration. For OCT500, the model accuracy rises from 0.8788 (GT) to 0.9697 (TR), recall from 0.7143 to 0.9286 and we can observe an improvement in F1 and AUC as well. For UIC, accuracy with TR OCTA is comparable to GT OCTA while recall remains near-identical. The combined dataset shows the best performance for TR OCTA, depicting a balanced dataset would perform well if the translation process generates good quality dataset.\\
The reason behind this is because GT OCTA enface images are subject to motion artifacts and scanner-to-scanner variability in noise characteristics and projection depth ranges. TR OCTA, by contrast, is generated by a model that has learned the stable structural-to-vascular correspondences that hold across patients. Because these correspondences are learned statistically from many patients, the translated image is effectively a denoised, average-representative vascular prediction that emphasizes exactly the disease specific features while suppressing random acquisition noise. When this cleaner signal is passed into the fusion branch, the cross-modal attention operates on a feature space that is more directly aligned with pathological meaning, which is why recall in particular improves so remarkably. The model is no longer confused by noisy negative OCTA regions that happen to resemble pathological features in GT images from some patients.\\
We understand that within-dataset performance is necessary, however not sufficient conditions for clinical deployment. So, we also designed cross dataset experiments since they are the most diagnostically relevant results in this study. When the OCT500-trained model is tested on UIC, the unimodal baseline (Table I) collapses to an accuracy of 0.5735 and an F1-score of 0.5396, barely above chance. The model has clearly learned OCT500 specific feature pattern rather than generalizable DR pathology. Crucially, its AUC of 0.9003 remains high, depicting the model’s ability to rank DR cases above Normal in a relative sense. The GT OCTA fusion model, on the other hand recovers substantially as accuracy and AUC improves in this configuration. The attention module’s reliance on vascular features that are more biologically is the likely mechanism since OCT B-scan texture is notoriously scanner dependent. Nevertheless, recall in this cross-dataset direction drops suggesting that while the model identifies DR cases correctly ranked, many DR cases are classified as false negative which could be contributed to the varying quality of used UIC DR dataset.\\
The TR OCTA fusion model in the same OCT500→UIC direction achieves both high accuracy and high recall, a combination that is clinically meaningful and that the GT variant cannot simultaneously achieve. The UIC→OCT500 direction remains the hardest scenario for all models, with TR OCTA yielding lower accuracy but higher AUC than GT OCTA and far better than the unimodal baseline. This residual gap reflects the genuine asymmetry between the two datasets: UIC contains predominantly moderate-to-severe DR cases, whereas OCT500 includes milder cases from a research cohort. A model trained on severe UIC pathology does not encounter the mild, structurally subtle DR presentations in OCT500 during training, leading to under performance of these borderline cases at test time.\\
Although we got good results for both GT OCTA and TR OCTA as expected, several limitations constrain the generalizability of the current findings. The dataset sizes remain modest: only 64 DR and 251 Normal patients in OCT500, and 415 patients in UIC limiting the diversity of disease presentations seen during training. Most publicly available OCT datasets are limited in size and phenotypic diversity that create bottleneck for generalization. The current study addresses only binary DR-versus-NORMAL classification; extending to DR severity grading, which is the clinically actionable task for treatment decisions, would require substantially more labelled data at each severity level. Moreover, the translation algorithm is still far from being clinically applicable hence there is much scope for improvement for the TR OCTA performance in this sector. Different dataset having different scanners and acquisition protocol along with different population also affect the model performance specially the UIC dataset since its quality is inconsistent unlike OCT500. \\
The primary clinical significance of this work is not just the performance numbers, but the pipeline architecture they validate: a system that requires only a standard OCT scanner at the point of care, uses a learned translation to synthesize vascular information and applies attention-based fusion to produce a classification output that approaches the performance of a system with full OCTA hardware. 
Diabetic retinopathy affects millions of people globally, with projections suggesting this will grow more as diabetes prevalence continues to rise, disproportionately burdening low- and middle-income regions where specialist ophthalmology access is scarce. Since OCTA devices remain expensive, requiring a high-end add-on to an already costly OCT system, and are absent from the vast majority of primary care and screening settings where the bulk of DR surveillance occurs. By demonstrating that TR OCTA fusion achieves substantially higher sensitivity than OCT this work shows that the hardware gap between well-resourced specialist clinics and underserved screening environments can be partly bridged through computation. \\
In summary, this work demonstrates that cross-modal attention fusion of OCT and OCTA enface images provides a substantial and consistent improvement over single-modal OCT classification for diabetic retinopathy detection. The TR OCTA variant, which eliminates the need for physical OCTA hardware achieves better performance across most of the experimental conditions including cross-dataset scenarios and matches or exceeds the performance reported by the recent studies. Together, these findings validate cross-modal OCT–OCTA enface fusion as a clinically meaningful paradigm for automated DR screening and the TR OCTA pathway in particular as an alternative route toward accessible, high-sensitivity retinal disease detection.

\section{Funding}
Supported by NEI R21EY035271 (MNA), R15EY035804 (MNA), 1R01EY037828-01 (MNA) and UNC Charlotte Faculty Research Grant (MNA).



\newpage
\begin{thebibliography}{99}

\bibitem{huang1991optical}  
D.~Huang \emph{et al.}, ``Optical coherence tomography,'' \emph{Science}, vol.~254, no.~5035, pp.~1178--1181, 1991.

\bibitem{fercher2003optical} 
A.~F. Fercher \emph{et al.}, ``Optical coherence tomography---principles and applications,'' \emph{Rep. Prog. Phys.}, vol.~66, no.~2, p.~239, 2003.

\bibitem{drexler2008optical} 
W.~Drexler and J.~G. Fujimoto, ``Optical coherence tomography: technology and applications,'' \emph{Springer}, 2008.

\bibitem{schmidt2018retina} 
E.~Schmidt-Erfurth \emph{et al.}, ``Artificial intelligence in retina,'' \emph{Prog. Retin. Eye Res.}, vol.~67, pp.~1--29, 2018.

\bibitem{wang2017optical} 
R.~K. Wang \emph{et al.}, ``Optical coherence tomography angiography-based capillary velocimetry,'' \emph{J. Biomed. Opt.}, vol.~22, no.~6, p.~066008, 2017.

\bibitem{gao2016quantitative} 
S.~S. Gao \emph{et al.}, ``Quantitative optical coherence tomography angiography,'' \emph{Biomed. Opt. Express}, vol.~7, no.~2, pp.~323--346, 2016.

\bibitem{spaide2018optical} 
R.~F. Spaide \emph{et al.}, ``Optical coherence tomography angiography,'' \emph{Prog. Retin. Eye Res.}, vol.~64, pp.~1--55, 2018.

\bibitem{an2016quantitative} 
L.~An \emph{et al.}, ``Quantitative comparisons between optical coherence tomography angiography and matched histology in the human eye,'' \emph{Exp. Eye Res.}, vol.~146, pp.~10--17, 2016.

\bibitem{ran2019development} 
A.~R. Ran \emph{et al.}, ``Development of a deep learning algorithm for optical coherence tomography angiography in diabetic retinopathy,'' \emph{Graefes Arch. Clin. Exp. Ophthalmol.}, vol.~257, pp.~1353--1361, 2019.

\bibitem{li2019deep} 
Z.~Li \emph{et al.}, ``Deep learning for detecting retinal detachment and discerning macular status using ultra-widefield fundus images,'' \emph{Commun. Biol.}, vol.~3, no.~1, p.~15, 2020.

\bibitem{Srinivasan2021} 
P.~P. Srinivasan \emph{et al.}, ``Fully automated detection of diabetic macular edema and dry age-related macular degeneration from optical coherence tomography images,'' \emph{Biomed. Opt. Express}, vol.~5, no.~10, pp.~3568--3577, 2014.

\bibitem{diffusion_mine}  
R.~H.~Badhon \emph{et al.}, ``Diffusion model based OCT to OCTA translation,'' \emph{Frontiers in Medicine}, vol.~12, 2025, doi: 10.3389/fmed.2025.1655453.

\bibitem{zhang2020multimodal}  
Q.~Zhang \emph{et al.}, ``Multimodal deep learning for macular disease detection,'' \emph{IEEE Trans. Med. Imaging}, vol.~39, no.~12, pp.~4093--4104, 2020.

\bibitem{zhou2021joint} 
Y.~Zhou \emph{et al.}, ``Joint classification and segmentation of diabetic retinopathy using deep learning,'' \emph{IEEE J. Biomed. Health Inform.}, vol.~25, no.~7, pp.~2671--2679, 2021.

\bibitem{wu2022cross}   
J.~Wu \emph{et al.}, ``Cross-modal learning for multi-modal medical image analysis,'' \emph{Med. Image Anal.}, vol.~78, p.~102394, 2022.

\bibitem{li2024review}  
Y.~Li \emph{et al.}, ``A review of deep learning-based information fusion techniques for multimodal medical image classification,'' \emph{Comput. Biol. Med.}, vol.~177, p.~108635, 2024.

\bibitem{azam2022review}    
M.~A. Azam \emph{et al.}, ``A review on multimodal medical image fusion: Compendious analysis of medical modalities, multimodal databases, fusion techniques and quality metrics,'' \emph{Comput. Biol. Med.}, vol.~144, p.~105253, 2022.

\bibitem{haribabu2023recent}  
M.~Haribabu, V.~Guruviah, and P.~Yogarajah, ``Recent advancements in multimodal medical image fusion techniques for better diagnosis: an overview,'' \emph{Curr. Med. Imaging}, vol.~19, pp.~673--694, 2023.

\bibitem{pei2023review}  
X.~Pei \emph{et al.}, ``A review of the application of multi-modal deep learning in medicine: bibliometrics and future directions,'' \emph{Int. J. Comput. Intell. Syst.}, vol.~16, p.~44, 2023.

\bibitem{ryu2023optimizing}  
G.~Ryu \emph{et al.}, ``Optimizing the OCTA layer fusion option for deep learning classification of diabetic retinopathy,'' \emph{Biomed. Opt. Express}, vol.~14, no.~9, pp.~4706--4722, 2023.

\bibitem{he2021multimodal}  
X.~He \emph{et al.}, ``Multi-modal retinal image classification with modality-specific attention network,'' \emph{IEEE Trans. Med. Imaging}, vol.~40, no.~6, pp.~1591--1602, 2021.

\bibitem{heisler2020ensemble}  
M.~Heisler \emph{et al.}, ``Ensemble deep learning for diabetic retinopathy detection using optical coherence tomography angiography,'' \emph{Transl. Vis. Sci. Technol.}, vol.~9, no.~2, p.~20, 2020.

\bibitem{huang2024comprehensive}  
Y.~Huang \emph{et al.}, ``A comprehensive review of multimodal deep learning for enhanced medical diagnostics,'' \emph{Clin. eHealth}, 2025, in press.

\bibitem{sun2024cmaf}  
K.~Sun \emph{et al.}, ``CMAF-Net: a cross-modal attention fusion-based deep neural network for incomplete multi-modal brain tumor segmentation,'' \emph{Quant. Imaging Med. Surg.}, vol.~14, no.~7, pp.~4579--4604, 2024.

\bibitem{li2024crossfuse}  
H.~Li and X.-J. Wu, ``CrossFuse: A novel cross attention mechanism based infrared and visible image fusion approach,'' \emph{Inf. Fusion}, vol.~103, p.~102147, 2024.

\bibitem{qu2022transmef}  
L.~Qu \emph{et al.}, ``TransMEF: A transformer-based multi-exposure image fusion framework using self-supervised multi-task learning,'' in \emph{Proc. AAAI Conf. Artif. Intell.}, vol.~36, 2022, pp.~2126--2134.

\bibitem{tang2022matr}  
W.~Tang \emph{et al.}, ``MATR: Multimodal medical image fusion via multiscale adaptive transformer,'' \emph{IEEE Trans. Image Process.}, vol.~31, pp.~5134--5149, 2022.

\bibitem{wang2024mactfusion}  
Z.~Wang \emph{et al.}, ``MACTFusion: Lightweight cross transformer for adaptive multimodal medical image fusion,'' \emph{J. King Saud Univ. Comput. Inf. Sci.}, vol.~36, no.~4, p.~102039, 2024.
  
\bibitem{yang2024ecfusion}  
X.~Yang \emph{et al.}, ``A novel multimodel medical image fusion framework with edge enhancement and cross-scale transformer,'' \emph{Sci. Rep.}, vol.~15, p.~6245, 2025.

\bibitem{liu2024multivit}  
Y.~Liu \emph{et al.}, ``A multimodal vision transformer for interpretable fusion of functional and structural neuroimaging data,'' \emph{Hum. Brain Mapp.}, vol.~45, p.~e26850, 2024.

\bibitem{alam2020avnet}       
M.~Alam \emph{et al.}, ``AV-Net: deep learning for fully automated artery-vein classification in optical coherence tomography angiography,'' \emph{Biomed. Opt. Express}, vol.~11, no.~9, pp.~5249--5257, 2020.



\bibitem{abtahi2022mfavnet} 
M.~Abtahi \emph{et al.}, ``MF-AV-Net: an open-source deep learning network with multimodal fusion options for artery-vein segmentation in OCT angiography,'' \emph{Biomed. Opt. Express}, vol.~13, no.~9, pp.~4870--4888, 2022.

\bibitem{liu2023hybrid}  
Y.~Liu \emph{et al.}, ``Hybrid fusion of high-resolution and ultra-widefield OCTA acquisitions for the automatic diagnosis of diabetic retinopathy,'' \emph{Diagnostics}, vol.~13, p.~2881, 2023.

\bibitem{Hongyi Pan 2025}  
H.~Pan \emph{et al.}, ``Multi-modal classification of retinal disease based on convolutional neural network,'' \emph{Biomedical Physics \& Engineering Express}, vol.~11, no.~4, 2025, doi: 10.1088/2057-1976/adeb92.

\bibitem{liu2024multieye}  
Z.~Liu \emph{et al.}, ``MultiEYE: Dataset and benchmark for OCT-enhanced retinal disease recognition from fundus images,'' arXiv preprint arXiv:2412.09402, 2024.

\bibitem{OCT500dataset}   
M.~Li \emph{et al.}, ``OCTA-500: A retinal dataset for optical coherence tomography angiography study,'' \emph{Medical Image Analysis}, vol.~93, p.~103092, 2024. :contentReference[oaicite:0]{index=0}

\bibitem{convnextv2}  
S.~Woo \emph{et al.}, ``ConvNeXt V2: Co-designing and Scaling ConvNets with Masked Autoencoders,'' arXiv preprint arXiv:2301.00808, 2023.

\bibitem{mainreference}  
Z.~Liu \emph{et al.}, ``Cross-modal attention network for retinal disease classification based on multi-modal images,'' \emph{Biomedical Optics Express}, vol.~15, no.~6, pp.~3699--3714, 2024.

\end{thebibliography}
\end{document}